# Tell me Mr. AI, what do you see in this image?

Tommaso Giacometti [1, 2, *], Paola Surcinelli [3, *], Mariachiara Stellato [1, 2, †], Nico Curti [1, 2]

[1] Department of Physics and Astronomy, University of Bologna, 40127 Bologna (Italy)
[2] INFN Bologna (Italy)
[3] Department of Psychology "Renzo Canestrari", University of Bologna, 40126 Bologna (Italy)
[†] *Corresponding Author:* M. Stellato (m.stellato@unibo.it)
[*] These authors contributed equally to this work

**Author Information**
- **Dr. Tommaso Giacometti** (0009-0002-5734-4579) tommaso.giacometti5@unibo.it
- **Prof. Paola Surcinelli:** (0000-0001-7875-3545) paola.surcinelli3@unibo.it
- **Dr. Mariachiara Stellato:** (0000-0003-4425-4427) m.stellato@unibo.it
- **PhD. Nico Curti:** (0000-0001-5802-1195) nico.curti2@unibo.it

**Statement and declarations**

**Funding**
This research received no external funding and was conducted with institutional support.

**Ethical statement**
Not applicable.

**Data sharing**
The dataset used for the training of the models is publicly available on the original ImageNet website (https://www.image-net.org/). The Rorschach image set is publicly available online on the Wikipedia website (https://it.wikipedia.org/wiki/Test_di_Rorschach). The code developed for the classification and statistical analysis of the results is publicly available on GitHub at https://github.com/Nico-Curti/RorschachAI.

**Acknowledgements**
The authors acknowledge the INFN CSN5 AIM-MIA Research Project.

**Declaration of interest**
The authors declare no competing interests.

**Contributors**
NC: data curation, investigation, visualization, conceptualization, supervision, methodology, formal analysis, and writing—original draft. MS: data curation, investigation, methodology and writing—review and editing. TG: data curation, investigation, methodology and writing—review and editing. PS: investigation, conceptualization, supervision, methodology, resources, project administration and writing—original draft.

## Abstract

**Background:** The Rorschach inkblots are ambiguous stimuli developed to evoke subjective interpretations in humans, while modern artificial intelligence (AI) models are trained to recognize well established patterns and classes. The comparison of these two opposite systems arises a simple and provocative question: what happens when we ask an AI model to interpret an inkblot that "is not supposed to represent anything predefined"?

**Methods:** We submitted the complete set of ten Rorschach inkblots to 61 AI models pretrained on the ImageNet dataset, spanning multiple architectural families. Model predictions were analyzed at the level of top-ranked classes and were quantified using a selected set of psycho-semantic variables inspired by the Rorschach tradition. Statistical analyses examined the effects of model family, computational complexity, and image conditions, comparing model-generated responses with human reference profiles.

**Findings:** Across all architectures, model responses were highly non-random and showed systematic semantic convergence and inter-model agreement. However, quantitative analyses revealed a clear and robust separation between human responses and all AI model families. Human profiles exhibited substantially higher affective load, semantic richness, projected agency, and variability, whereas AI models converged toward frequent, formally coherent, and perceptually stable interpretations.

**Interpretation:** Vision models consistently project the semantic organization learned, favoring consensus and formal coherence over affective or symbolic elaboration. Applying the Rorschach test to AI systems does not assess human-like cognition but provides a principled framework for exposing perceptual and semantic biases embedded in contemporary computer vision models.

## 1. Introduction

The Rorschach inkblots were born to confound. They do not represent any defined and/or specific object, they do not suggest a correct answer, and they do not offer a stable ground truth of interpretation (Gelso & Williams, 2022; Vitolo et al., 2021; Weiner, 2021). Their value, according to traditional psychology, relies on their ambiguity: what is seen does not provide so much about the image, but on the observer (Campo, 2017; Choca & Rossini, 2018). The ambiguity of the Rorschach inkblots aims to evoke individualized responses. This lack of inherent structure encourages subjects to project their own perceptions, emotions, and cognitive patterns, thereby providing insights into personality organization, underlying affective states, and thought processes. In contrast, the modern artificial intelligence (AI) models are trained to minimize ambiguity, mapping features and patterns learnt on a limited amount of well-established semantic classes (Übellacker, 2025). Ask to an AI model "What do you see?" on a Rorschach picture means to undermine the solidity of its training, designedly putting it outside its comfort zone.

This work was born from a question as simple as uncommon: what happens to a system designed to recognize dogs, airplanes and teapots when we provide a picture that does not represent anything? The purpose is not to evaluate the "correctness" of the AI answers (that is senseless in this context), but to observe how different AI models, based on their complexity (Hu et al., 2021), computational architecture (Liang, 2025), and types (Horvath & Pouliou, 2024; Vernon, 1935) could respond to a stimulus that does not contain any object (ref. **Figure 1**).

We analyzed the predictions of a wide set of pre-trained AI models using the ImageNet dataset (Deng et al., 2009). The selection of the models was performed considering the most common and used architectures (Albukhari, 2025; He et al., 2015; Li et al., 2025; Simonyan & Zisserman, 2014; Tan & Le, 2019) in the AI research field. It is important to note that most of the modern AI applications, indeed, typically used one or more of these models as backbone of their pipeline (Albukhari, 2025; Li et al., 2025), providing only fine-tuning of the model parameters. According to the AI community, the ImageNet dataset represents the gold standard of classification task, providing a robust starting point for model re-training (Florence et al., 2025) or the so-called transfer learning procedure (Zhuang et al., 2019). The heterogeneity of pictures included in this dataset, indeed, could guarantee a quasi-optimal initial condition (Vasey et al., 2025), including textures, shapes, and patterns commonly found in other (apparently far) applications (Devireddy, 2025; Selvaraju et al., 2020).

From a psychological perspective, this design offers a distinctive research opportunity (Areh et al., 2022; Kimoto et al., 2017; Petot & Djurić Jočić, 2005; Vecchio et al., 2023): it allows the examination of responses to Rorschach stimuli in the absence of emotional, affective, and motivational processes that inevitably influence human performance. By applying the same ambiguous material to different AI models under fully controlled conditions, it becomes possible to isolate interpretative patterns driven solely by representational and associative structures derived from training data. This approach enables systematic analysis of how meaning is generated from ambiguity independently of emotional reactivity, defense mechanisms, or subjective experience, which are central in human clinical assessment.

The analysis of models' predictions, stratifying them according to their complexity and model "families" (He et al., 2015; Simonyan & Zisserman, 2014; Tan & Le, 2019), allow to measure the presence of shared pattern (Albukhari, 2025; Li et al., 2025), semantic convergences (Son et al., 2025) or recurrent perceptive strategies (Ji-An et al., 2025). The attention of our analysis was not focused on the single response ("this image contains a bat"), but to the global structure of the answers: models' agreement, conceptual distance among predicted classes (Deng et al., 2009) and image regions which guide the decision-making process (Selvaraju et al., 2020), exactly in the same way of a human-like test.

The main strength of this work relies on the changing of perspective: pictures designed to test human subjectivity are used as methods to investigate perceptive, semantic, and architectural biases (Hanna et al., 2025; Laurito et al., 2025; Zhou et al., 2024) in the most common AI models. While AI systems can simulate projective interpretative processes, the worldview they express is fundamentally determined by their training data. Therefore, their answers to the Rorschach test could represent an overview of the conceptual space of the ImageNet dataset and patterns learnable by models trained on it. This approach offers also a novel instrument to compare AI models beyond standard metrics, focusing on their coherence, agreement, and internal representation structure.

In conclusion, to ask to an AI model to interpret the Rorschach test does not aim to understand if it "thinks like a human", but to better appreciate the presence of biases and possible pitfalls in the model training which may be projected into real-world applications. AI-specific inkblot tests would not measure personality, emotion, or unconscious conflict. Instead, it would function as a diagnostic probe of model structure, training, and inference behavior under ambiguity. Any apparent similarity is surface-level and statistical, not diagnostic.

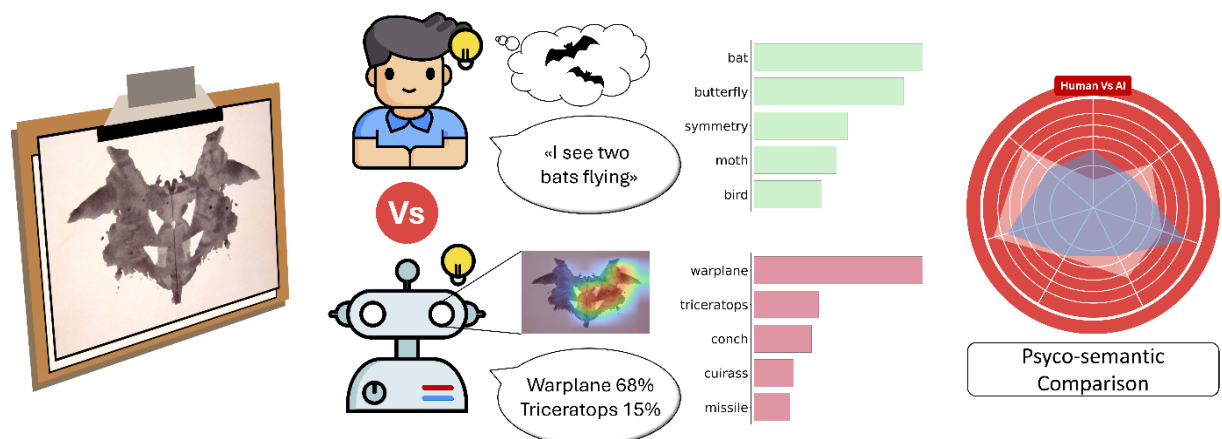

**Figure 1 | Schematic representation of the proposed pipeline.** Our analysis is focused on the quantitative comparison between Rorschach test responses proposed by a wide range of AI models and standard human outcomes; by asking to an AI model to interpret the Rorschach test (like the one in reported in the picture), we aim to compare the psyco-semantic characteristics of the responses, focusing on a better understanding about the presence of possible biases and pitfalls in the AI model training. The responses and frequencies reported in the picture are related to the inkblot shown on the left.

## 2. Materials and Methods

*2.1 Rorschach inkblots digitization*

In this work, we used the full set of 10 digitized Rorschach inkblots (Vernon, 1935), publicly available online. The images were resized and normalized according to the requirements of the different AI models, preserving their spatial structure, bilateral symmetries, and contrast. No other pre-processing was applied, aiming to keep the intrinsic ambiguity of the stimuli. We would like to point out that 5/10 images are intrinsically greyscale, while the remaining ones have at least one color. In the human Rorschach test, achromatic stimuli are primarily associated with cognitive and perceptual processing, red elements with intense affective arousal, and chromatic colors with broader emotional activation and affect modulation: to guarantee a correct harmonization of all the images, during the evaluation of the models all the pictures were considered as colored, eventually triplicating the grayscale channel.

*2.2 Artificial Intelligence models*

A total of 61 AI vision models were analyzed and implemented using the PyTorch framework (Paszke et al., 2019). All models were instantiated with publicly available pre-trained weights obtained from the ImageNet-1k dataset (Deng et al., 2009) (1000 annotated classes), accessed on December 18, 2025. The selection was designed to cover a broad range of architectural families, including convolutional neural networks (O'Shea & Nash, 2015), residual networks (He et al., 2015), and efficient architectures (Tan & Le, 2019).

All models shared the same output label space (ImageNet classes), allowing direct comparison of predictions without any post-hoc relabeling or semantic alignment. For each model, additional metadata was collected, including *model family* (i.e. the category of model architecture), *model name*, *number of parameters*, and estimated computational cost in *BFLOPs* (Billion FLoating point Operations Per Second). The complete list of models and their characteristics is provided in the *Supplementary Material*.

Each model was evaluated in inference-only mode, without any further refinement, i.e. keeping all parameters frozen and without any task-specific fine-tuning or retraining.

*2.3 Model inference and prediction extraction*

Each model was presented with the full set of Rorschach inkblots. For each image-model pair, we extracted the top five predicted classes along with their associated confidence score. Predictions were computed for multiple experimental conditions, including different image augmentations (Perez & Wang, 2017) (original image, horizontal flip, vertical flip), which were tracked explicitly in the dataset. All subsequent analyses were performed at the level of individual image-model predictions, preserving the full structure of repeated observations across images, models, and augmentation conditions.

*2.4 Psycho-Semantic Analysis*

To enable a direct comparison between human and AI Rorschach responses, each predicted class label was associated with a set of psycho-semantic scores. In the context of the Rorschach test, "popular responses" are defined as interpretations that occur with statistically significant frequency in human populations and are linked to shared perceptual configurations (Molish, 1951). Beyond traditional content codes, responses can be analyzed according to higher-order semantic dimensions. Following this framework, both human responses reported in the literature (ref. *Supplementary Materials*) and

AI predictions were mapped onto an *ad hoc* and manually curated set of eleven psycho-semantic scores.

The *percentage frequency* of each response was computed as $f_i = \frac{n_i}{N} \times 100$, where $n_i$ denotes the number of occurrences of a given response or semantic category and $N$ is the total number of responses, considering both human data from the literature and the full set of model predictions.
*Emotional valence* captures the intrinsic affective polarity of the semantic content of the response, while *affective load* was defined as emotional salience beyond neutral description. *Aggressiveness* and *anxiety*-related content were computed as explicit or implicit violent or offensive content and potential of the response content to elicit fear, danger, or aversion, respectively. *Perceptual adaptivity* and *formal integration* were quantified as the degree of semantic conventionality of the response and their internal semantic coherence. *Social orientation* estimated whether the semantic content implies interpersonal interaction, while *human presence* quantifies the explicit or implicit reference to the human body or human agents. *Semantic complexity* reflects the structural richness of the response, while *dynamicity* reflects the proportion of predictions associated with actions or motion-related concepts.

All indices were computed independently for each $(image, model, augmentation)$ instance and normalized to ensure comparability across variables. A detailed mathematical description of the psycho-semantic indices is provided in the *Supplementary Materials*.

## *2.5 Statistical Analysis*

Statistical analyses were designed to assess how psycho-semantic variables depend on model-related and experimental factors. Specifically, we examined the effects of *model family*, *augmentation*, *semantic domain* of the response, *number of model parameters*, and *BLOPs* (as proxy of the execution time of the evaluation).

Descriptive statistics were computed for all psycho-semantic variables, stratified by *model family*, including a reference human condition. Group-level differences were assessed using linear models and analysis of variance, treating categorical variables (*model family, augmentation, semantic domain*) as fixed effects. Continuous associations with *number of parameters* and *BFLOPs* were evaluated using non-parametric correlation analyses to account for non-normal distributions.
For each Rorschach inkblot we generated an independent embedding space, i.e. a projection of the internal model representation for visual inspection, on which each point represents a model and the relative distance between points represents the similarity between models' activation patterns.

## 3. Results

We performed the statistical analysis using the 11 psycho-semantic scores, monitoring their effect on the different types of AI-/human- models. We found a systematic effect on model types, proving a significant difference in human responses compared to the AI counterparts. The answers proposed by humans showed significant average differences (higher) in variables linked to affection, projection, and semantic complexity (*affective load*, *human presence*, *anxiety*, *aggressiveness*, *semantic complexity*, *dynamicity*), while the AI models are significantly stronger (higher values with low variances) on perceptual-formal dimensions (*formal integration*, *perceptive adaptivity*) (ref. **Table 1**). The inter-variability of AI model types appeared more homogeneous, relying on a more compact psychological space, well split to human portion (ref. **Figure 2**).

**Table 1 | Human versus AI responses and main effect of model family.** Descriptive statistics (mean ± SD) for human and AI responses across psycho-semantic variables. Statistical significance refers to the fixed effect of the human condition estimated using linear regression models on all observations, i.e. variable ~ is_human + (1 | image).

| Psycho-semantic Metric | Human (Mean ± SD) | AI (Mean ± SD) | Human vs AI trend | β (Human) | p-value * |
|---|---|---|---|---|---|
| Emotional valence | 0.14 ± 1.14 | -0.15 ± 0.93 | Human > AI | + | < 0.05 * |
| Affective load | 1.50 ± 0.76 | 0.32 ± 0.47 | Human >> AI | + | < 0.001 *** |
| Perceptual adaptivity | 0.71 ± 0.18 | 0.84 ± 0.09 | AI > Human | - | < 0.01 ** |
| Formal integration | 1.64 ± 0.48 | 1.92 ± 0.22 | AI > Human | - | < 0.001 *** |
| Social orientation | 0.08 ± 0.44 | 0.03 ± 0.17 | Human > AI | + | n.s. |
| Human presence | 0.52 ± 0.81 | 0.04 ± 0.20 | Human >> AI | + | < 0.001 *** |
| Aggressiveness | 0.44 ± 0.86 | 0.08 ± 0.25 | Human > AI | + | < 0.001 *** |
| Anxiety / threat | 0.78 ± 0.95 | 0.12 ± 0.31 | Human >> AI | + | < 0.001 *** |
| Semantic complexity | 1.46 ± 0.68 | 1.03 ± 0.26 | Human > AI | + | < 0.001 *** |
| Dynamicity | 0.54 ± 0.68 | 0.30 ± 0.46 | Human > AI | + | < 0.01 ** |
| Response frequency (%) | Broad distribution | High, concentrated | AI > Human | - | < 0.001 *** |

*\* p-values refer to the fixed effect of the human condition estimated using linear models on individual observations. This approach avoids pseudo-replication and accounts for the unbalanced structure of the data. Asterisks indicate the significance level of the main effect of model family (\*\*\* p < .001, \*\* p < .01, \* p < .05, n.s. not significant).*

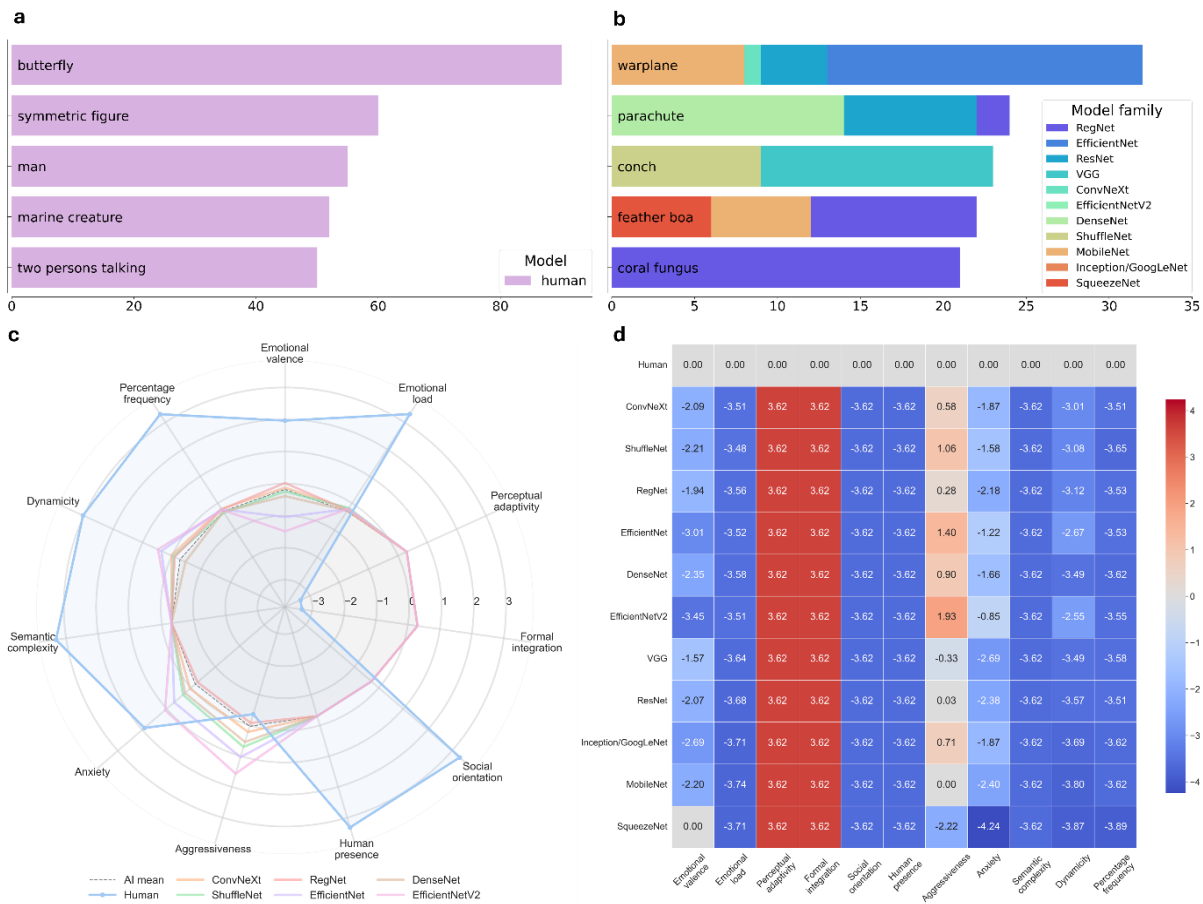

**Figure 2 | Comparison of psycho-semantic profiles between AI models and human. a.** Most frequent prediction for human patients. **b.** Most frequent predictions for the AI models, stratified by model family (top 5 per family). Despite the absence of ground truth, models converge on a limited set of recurrent semantic categories, revealing architecture-dependent interpretative biases. **c.** Contributions of psycho-semantic metrics in the description of the model behaviors; for each metric, the standardized (z-score) contribution is reported in the radar-plot representation, highlighting the 'human' model with the blue line. **d.** Relative percentage of the psycho-semantic metrics (z-score) using the human outcomes as baseline.

The variance analysis showed no statistical effect on the psychological variables considered in relation to the image orientation, i.e. the opportunity of analyzing the image or any kind of flipped version does not appear correlated with the obtained outcome. In contrast, the semantic domain of the answers seems to be moderately linked to some structural dimensions and percentage frequencies of the responses. The parameters describing the model complexity (*number of parameters* and *BFLOPs*) showed low but significant correlations with the reduction of conflictual contents (*aggressivity*, *anxiety*), without explaining the gap between AI and human responses.

The multivariate analysis of the distance between AI and humans, based on the standardized average profiles, provides the opportunity to estimate a ranking of the model types based on their similarity to humans. Nevertheless, the intra variability of AI model types appeared limited if compared to the overall gap between them and humans. The main contribution of this distance could be imputed to the strong effectiveness of affective variables and projection components which characterize human answers.

The projection of the model predictions in the estimated embedding space (ref. *Supplementary Materials* for details), split according to the different Rorschach inkblots, is reported in **Figure 3**. The points were colored according to the analyzed model families, allowing a visual inspection of possible clusters inter- and intra- model families. This analysis shows the presence of recurrent patterns in the projected spaces, highlighting how architectures belonging to the same families tend to occupy contiguous regions in the embedding space. These results support the hypothesis that models' responses are not random (despite the missing semantic "ground truth" of the Rorschach inkblots) but well-structured according to the different architectural features, proving the presence of biases on the different model families.

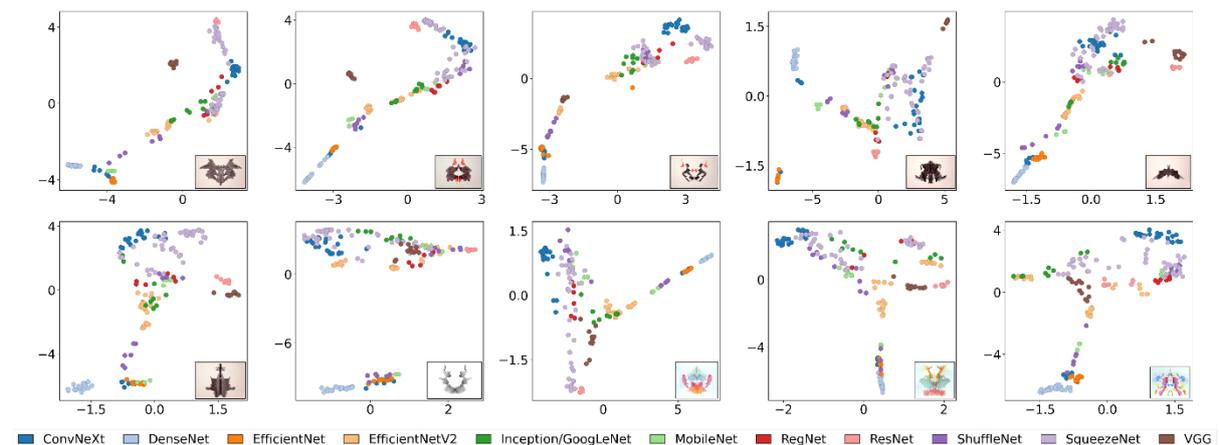

**Figure 3 | Embedding spaces of model response similarities across Rorschach inkblots.** For each Rorschach image, we report the PaCMAP *(Chatzidimitris et al., 2015)* projection of the standardized logit vectors produced by all models. Each point represents a single model, and distances between points reflect similarity in their internal response patterns to the same inkblot. Points are colored according to model architectural family. For each scatter plot we reported also the respective inkblot.

## 4. Discussion

The obtained results provide a quantitative and systematic answer to the initial question motivating this work: "*what happens when artificial vision systems, explicitly designed to minimize ambiguity, are exposed to stimuli whose psychological function is precisely to maximize it?*" The analyses consistently show that the Rorschach inkblots elicit responses from AI models that are structurally coherent and highly constrained, yet fundamentally different from those produced by human observers.

Across all psycho-semantic variables, the most robust finding is the sharp separation between human interpretations and all families of AI models. This separation emerges clearly at the descriptive level and is confirmed by inferential analyses, with the main effect of *model family* being highly significant for most variables (typically p < 0.001) and largely driven by the contrast between human responses and model predictions. Human responses exhibit substantially higher *affective load* (mean = 1.50 vs. AI ≈ 0.32), greater attribution of *human presence* (0.52 vs. ≈ 0.04), and stronger expressions of *anxiety* (0.78 vs. ≈ 0.12) and *aggressiveness* (0.44 vs. ≈ 0.08), all with large effect sizes. In contrast, AI models systematically show higher *formal integration* and *perceptual adaptivity* ($p < 0.001$), together with markedly reduced variability. According to these results, it is important to note that the AI models considered were pre-trained on a dataset which does not include any reference to human figures or any related behavior (ref. *human presence* score). This poses a strong limitation on the current study mainly linked to the impossibility to directly compare the results between human subjects and AI models. Moreover, we would like to stress that the considered AI model ensemble does not include any modern gen-AI architecture (e.g. ChatGPT, Gemini), shrinking the results simply to fit the inkblots to a pre-determined and finite set of classes, without any room for more flexible evaluations.

These results directly reflect the conceptual tension surmised in our initial question. The Rorschach test does not probe recognition accuracy but rather the organization of perception under uncertainty. Human observers respond to this uncertainty by projecting affective, symbolic, and agentive content, producing responses that are both semantically diverse and emotionally loaded. By contrast, AI models respond to the same ambiguity by collapsing it into well-formed, internally consistent categorizations drawn from a fixed semantic space. Quantitatively, this is reflected in the strong convergence of model responses toward high *percentage frequency* values, and in the limited dispersion of their psycho-semantic profiles.

Importantly, the comparison across model families reveals that this pattern is remarkably stable. Although a ranking of families based on multivariate distance from the human profile can be established, the range of distances among AI architectures is narrow (approximately 10.36 to 11.50 in standardized space, ref. *Supplementary Materials*), especially when contrasted with the absolute separation from the human condition. Even the architectures that appear relatively closer to humans (such as ConvNeXt or RegNet) remain far removed on the dimensions that most strongly characterize human Rorschach responses, including *affective load*, *human presence*, and *anxiety*-related content. This indicates that architectural diversity modulates how models organize ambiguity but does not fundamentally alter the nature of their responses.

Analyses involving computational scale further support this conclusion. Both the *number of parameters* and *BFLOPs* show statistically significant but weak correlations with *percentage frequency* (Spearman $\rho \approx 0.05$, $p < 10^{-6}$), indicating that larger models tend to converge more strongly toward frequent or shared responses. Smaller negative correlations are observed with *aggressiveness* and *anxiety* ($\rho \approx -0.02$ to $-0.03$, $p < 0.05$), suggesting a mild attenuation of conflictual content as scale increases. However, no significant relationship is found between model scale and the psycho-semantic dimensions that most clearly separate humans from models, such as *affective load* or attribution of agency. In other words, increasing model size does not increase ambiguity tolerance, but rather reinforces normative convergence within the learned semantic space.

The multivariate decomposition of distances makes this pattern explicit. As expected, across all model families, the largest contributors to the distance from the human profile are affective and projective variables, including *affective load*, *human presence*, *anxiety*, *aggressiveness*, and *response frequency*. Perceptual and formal variables contribute substantially less to the separation and, in some cases, show opposite trends, with models exceeding humans. This asymmetry clarifies that the human–AI gap is not primarily perceptual but psycho-semantic: models succeed in imposing structure on ambiguity, whereas humans use ambiguity as a space for projection.

Considering the surmised theoretical framing, these results suggest that applying the Rorschach test to AI models does not reveal whether they "think like humans" but rather expose the implicit constraints of their training and representational spaces. The responses generated by the models can be interpreted as projections of the ImageNet semantic universe under conditions of extreme under-specification. Rather than producing idiosyncratic or affectively rich interpretations, models consistently map ambiguous stimuli onto stable, high-frequency concepts, reflecting both the structure of the dataset and the optimization objectives underlying their training.

Taken together, the findings support the view that current computer vision systems implement a form of ambiguity reduction that is fundamentally different from human interpretive processes. While both humans and models may assign labels to the same stimulus, the psycho-semantic organization underlying those labels remains quantitatively and qualitatively distinct. From this perspective, the Rorschach inkblots function not as a test of artificial subjectivity, but as a methodological lens through which architectural, semantic, and training-related biases in AI models can be systematically revealed.

*4.1 Strengths*

A major strength of this work lies in the explicit shift of perspective: stimuli originally designed to probe human subjectivity are repurposed as instruments for the systematic analysis of artificial perceptual systems. This allows the investigation of AI models beyond standard performance metrics, focusing instead on coherence, convergence, and the structure of responses under ambiguity.

The study further benefits from the inclusion of a large and diverse set of pre-trained models spanning multiple architectural families and computational scales, all evaluated within a unified semantic output space. This design enables robust statistical comparisons and minimizes confounding factors related to task-specific fine-tuning. The combination of univariate, multivariate, and distance-based analyses provide converging evidence for the reported patterns, while the supplementary analyses of logit-space embeddings and Grad-CAM (Selvaraju et al., 2020) attention maps strengthen the link between output behavior and internal representations.

*4.2 Limits*

Several limitations should be considered. In principle, the analysis relies exclusively on models pre-trained on ImageNet, which constrains the semantic space available to the models and may limit the expression of alternative representational strategies.

Furthermore, the mapping between ImageNet labels and psycho-semantic variables necessarily abstracts complex psychological constructs and cannot fully capture the richness of human symbolic interpretation; it is also important to notice that ImageNet does not include any explicit human representation which, in contrast, play a key role in the common human responses.

This study used the Rorschach inkblots as experimental probes to investigate how current state-of-the-art AI models respond to extreme visual ambiguity. By analyzing a wide range of ImageNet-pretrained architectures through a psycho-semantic framework, we showed that human and model-generated responses occupy clearly distinct regions of the representational space.

Across all analyses, AI models converged toward stable, frequent, and formally coherent interpretations, largely independent of architectural family or computational scale, while human responses exhibited greater affective load, semantic richness, and variability. Differences among model families modulated response style but did not reduce the overall gap with the human profile.

Rather than revealing human-like interpretation, the Rorschach test exposes the structural constraints and biases embedded in AI models. As such, projective psychological stimuli provide a complementary and informative lens for comparing artificial systems beyond standard performance metrics.

The results of the study further underscore that, unlike its use in human clinical assessment, the application of the Rorschach test to artificial intelligence systems does not provide access to unconscious processes, personality structure, or emotional functioning. Analyses make explicit that the responses generated by the models do not reflect psychodynamic projective mechanisms but instead emerge as statistically grounded associations derived from patterns learned during training. In this context, the notion of "projection" is purely metaphorical and refers to the way in which training-related biases and representational constraints shape the interpretation of ambiguous stimuli. Consequently, the AI-based use of the Rorschach should not be understood as a psychological assessment of the model, but rather as an analytical tool for examining how characteristics of the training process influence model outputs under conditions of perceptual ambiguity.

# Supplementary Materials

## Model Inference Pipeline and Parameter Extraction

All analyses were conducted using a unified and fully automated inference pipeline implemented in PyTorch and torchvision python packages. A total of 61 convolutional neural network (CNN) architectures, spanning multiple architectural families, were evaluated under identical experimental conditions. All models were pre-trained on the ImageNet-1k dataset and used without any fine-tuning, in order to isolate the effect of architectural inductive biases and learned visual representations.

For each model, inference was performed on the complete set of Rorschach inkblot images using the default preprocessing pipeline associated with the corresponding pretrained weights (including resizing, cropping, normalization, and interpolation strategy). This ensured that each architecture was evaluated under the conditions recommended by its original training protocol, while preserving comparability at the level of model outputs.

To assess robustness with respect to basic geometric transformations, each image was processed under three deterministic variants: (i) the original image, (ii) a horizontally flipped version, and (iii) a vertically flipped version.

Each variant was explicitly tracked through an augmentation identifier in the final dataset, allowing downstream analyses to quantify invariance or sensitivity of model predictions to spatial transformations.

For every (model, image, augmentation) triplet, the full forward pass produced a vector of pre-softmax logits over the 1000 ImageNet classes. These logits were retained in full and stored as serialized vectors, enabling subsequent analyses based on distributional similarity, semantic embedding, and inter-model agreement. From the softmax-normalized probabilities, the top 5 predicted classes and their associated probabilities were extracted and recorded, yielding a structured representation of both dominant and secondary model responses.

In addition to prediction outputs, extensive model-level metadata were systematically collected. These included the architectural family (e.g., ResNet, EfficientNet, ConvNeXt), total number of learnable parameters, an indicator of whether the model belongs to a lightweight/mobile-oriented class, and the nominal input resolution inferred from the pretrained weights metadata. This information was used to contextualize prediction behavior with respect to architectural complexity and design goals.

Finally, a computational complexity estimate was associated with each model in terms of BFLOPs (billion floating-point operations) required for a single forward pass at the effective input resolution. FLOPs were estimated using established profiling tools when available, following the common convention of approximating FLOPs as twice the number of multiply–accumulate operations (MACs). When certain operators (e.g., pooling layers) were not supported by the profiler, the resulting BFLOPs were treated as approximate but remained suitable for relative comparisons across architectures.

All extracted quantities were aggregated into a single tabular dataset, ensuring full traceability of predictions, augmentations, model properties, and computational cost.

**Sup Tab 1 | Models' architecture and computational complexity.** Complete list of the 61 pretrained convolutional neural network architectures evaluated in this study, grouped by architectural family. For each model, the table reports the total number of trainable parameters (in millions) and the estimated computational complexity expressed as BFLOPs for a single

forward pass at the effective input resolution defined by the pretrained weights. BFLOPs were computed using automated profiling tools and should be interpreted as approximate but suitable for relative comparisons across architectures.

| Model family | Model name | Parameters (M) | BFLOPs |
|---|---|---|---|
| ResNet | resnet18 | 11.69 | 1.82 |
| | resnet34 | 21.80 | 3.67 |
| | resnet50 | 25.56 | 4.11 |
| | resnet101 | 44.55 | 7.83 |
| | resnet152 | 60.19 | 11.56 |
| | resnext50_32x4d | 25.03 | 4.26 |
| | resnext101_32x8d | 88.79 | 16.48 |
| | wide_resnet50_2 | 68.88 | 11.43 |
| | wide_resnet101_2 | 126.89 | 22.80 |
| VGG | vgg11 | 132.86 | 7.61 |
| | vgg11_bn | 132.87 | 7.62 |
| | vgg13 | 133.05 | 11.31 |
| | vgg13_bn | 133.05 | 11.33 |
| | vgg16 | 138.36 | 15.47 |
| | vgg16_bn | 138.37 | 15.50 |
| | vgg19 | 143.67 | 19.63 |
| | vgg19_bn | 143.68 | 19.66 |
| DenseNet | densenet121 | 7.98 | 2.87 |
| | densenet161 | 28.68 | 7.79 |
| | densenet169 | 14.15 | 3.40 |
| | densenet201 | 20.01 | 4.34 |
| EfficientNet | efficientnet_b0 | 5.29 | 0.40 |
| | efficientnet_b1 | 7.79 | 0.71 |
| | efficientnet_b2 | 9.11 | 1.13 |
| | efficientnet_b3 | 12.23 | 1.88 |
| | efficientnet_b4 | 19.34 | 4.51 |
| | efficientnet_b5 | 30.39 | 10.51 |
| | efficientnet_b6 | 43.04 | 19.47 |
| | efficientnet_b7 | 66.35 | 38.45 |
| EfficientNetV2 | efficientnet_v2_s | 21.46 | 8.45 |
| | efficientnet_v2_m | 54.14 | 24.79 |
| | efficientnet_v2_l | 118.52 | 56.44 |
| MobileNet | mobilenet_v2 | 3.50 | 0.31 |
| | mobilenet_v3_small | 2.54 | 0.06 |
| | mobilenet_v3_large | 5.48 | 0.23 |
| Inception / GoogLeNet | googlenet | 6.62 | 1.50 |
| | inception_v3 | 27.16 | 5.73 |
| SqueezeNet | squeezenet1_0 | 1.25 | 0.82 |
| | squeezenet1_1 | 1.24 | 0.35 |
| ShuffleNet | shufflenet_v2_x0_5 | 1.37 | 0.04 |
| | shufflenet_v2_x1_0 | 2.28 | 0.15 |
| | shufflenet_v2_x1_5 | 3.50 | 0.30 |
| | shufflenet_v2_x2_0 | 7.39 | 0.59 |
| RegNet | regnet_y_400mf | 4.34 | 0.41 |
| | regnet_y_800mf | 6.43 | 0.85 |
| | regnet_y_1_6gf | 11.20 | 1.63 |
| | regnet_y_3_2gf | 19.44 | 3.20 |
| | regnet_y_8gf | 39.38 | 8.52 |
| | regnet_y_16gf | 83.59 | 15.96 |
| | regnet_y_32gf | 145.05 | 32.35 |
| | regnet_x_400mf | 5.50 | 0.42 |

|  | regnet_x_800mf | 7.26 | 0.81 |
|---|---|---|---|
|  | regnet_x_1_6gf | 9.19 | 1.62 |
|  | regnet_x_3_2gf | 15.30 | 3.20 |
|  | regnet_x_8gf | 39.57 | 8.02 |
|  | regnet_x_16gf | 54.28 | 15.99 |
|  | regnet_x_32gf | 107.81 | 31.81 |
| **ConvNeXt** | convnext_tiny | 28.59 | 4.47 |
|  | convnext_small | 50.22 | 8.71 |
|  | convnext_base | 88.59 | 15.38 |
|  | convnext_large | 197.77 | 34.40 |

## Model responses analysis

To characterize the qualitative behavior of the AI models, we analyzed the most frequent class predictions produced by the models split according to the 10 Rorschach inkblots. For sake of clarity, the analysis was conducted considering only the top 1 prediction.

At the global level, the distribution of predicted classes shows a clear deviation from randomness. Despite the absence of a ground truth and the abstract nature of the stimuli, models consistently converged on a limited subset of ImageNet categories. The most frequent predictions include concrete objects and well-defined shapes such as *missile*, *parachute*, *warplane*, and *isopod*. These categories are characterized by strong structural priorities (elongated shapes, bilateral symmetry, central mass), suggesting that models tend to map the perceptual ambiguity of the inkblots onto familiar high-level visual templates learned during ImageNet pre-training.

The stratification across model families (ref. **Sup. Figure 1**) reveals systematic differences in semantic bias across architectural families. Convolutional and residual-based models (e.g., ResNet, RegNet, EfficientNet) tend to emphasize rigid, object-like interpretations (e.g., *missile*, *warplane*, *space shuttle*), while lightweight architectures (e.g., MobileNet, ShuffleNet, SqueezeNet) show a higher variability and a greater tendency toward texture- or part-based categories (e.g., *feather boa*, *paper towel*, *coral fungus*). In particular, VGG models display a strong concentration on a narrow set of recurring labels, indicating a more stereotyped response pattern.

Overall, this analysis highlights that, even in the absence of semantic constraints, AI models do not respond idiosyncratically to Rorschach inkblots. Instead, their interpretations are shaped by architectural inductive biases and shared training history, leading to reproducible and family-specific patterns of "projective" responses. These findings parallel classical observations in human Rorschach assessment, where perceptual organization and prior experience guide the attribution of meaning to ambiguous stimuli.

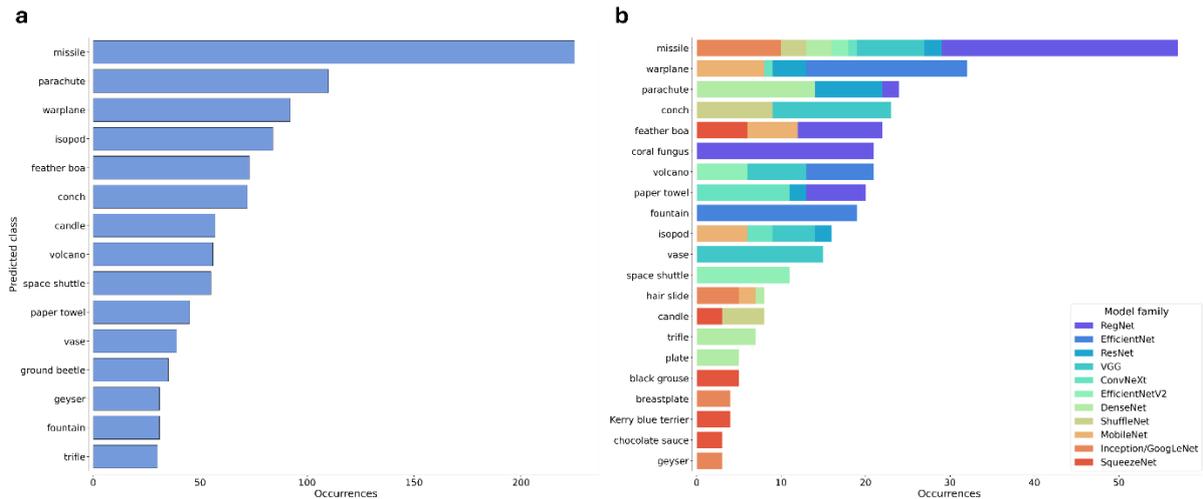

**Sup Figure 1 | Most frequent Top 1 predictions produced by AI models in response to Rorschach inkblots. a.** Global distribution of the 15 most common predicted classes across all models; **b.** most frequent predictions stratified by model family (top 5 per family). Despite the absence of ground truth, models converge on a limited set of recurrent semantic categories, revealing architecture-dependent interpretative biases.

To further characterize the nature of the model responses, predictions were analyzed by stratifying results across individual inkblots (and by grouping ImageNet classes into higher-level thematic and form-based categories (e.g., flying objects, animals, containers, natural formations, textiles). Only top 1 predictions were considered to capture the dominant perceptual interpretation of each model.

At the image level, each inkblot elicited a distinct and reproducible semantic profile. Some pictures showed a strong convergence toward a limited set of rigid and symmetric object categories (e.g., flying objects such as *missile* or *warplane*), while others exhibited a more heterogeneous thematic distribution including animals, soft materials, and natural forms. This pattern was quantified using normalized theme proportions and Shannon entropy. Inkblots characterized by elongated or bilateral structures displayed significantly lower thematic entropy, indicating stereotyped interpretations, whereas inkblots with fragmented or diffuse shapes produced higher entropy values, reflecting increased semantic dispersion.

A non-parametric Kruskal-Wallis test confirmed significant differences in thematic distributions across inkblots ($p \ll 0.01$), demonstrating that the models' predictions are systematically modulated by the structural properties of the stimuli rather than by random variability. Importantly, these effects were consistent across model families, suggesting that perceptual organization principles learned during ImageNet training dominate over architectural idiosyncrasies at this level of abstraction.

Overall, this analysis reveals that AI responses to Rorschach inkblots are structured along latent form- and theme-based dimensions, closely paralleling classical Rorschach constructs such as form dominance, content category, and perceptual organization.

**Sup Tab 2 | Quantitative thematic summary of top 1 prediction produced by all models for each Rorschach inkblot.** For each stimulus, the table reports the number of unique predicted labels, the frequency of labels split into perceptual and affective properties, and Shannon entropy of the thematic distribution.

| Inkblot | # unique classes | Threat (%) | Animacy (%) | Movement (%) | Soft (%) | Structure (%) | Natural (%) | Dimensional entropy (H) |
|---|---|---|---|---|---|---|---|---|
| I | 39 | **32.24** | 1.64 | 3.28 | 2.73 | 1.64 | 6.56 | 1.610 |
| II | 41 | 1.09 | 0.55 | 14.21 | **16.94** | 3.83 | 4.37 | 1.943 |
| III | 37 | 0.00 | 3.83 | 10.93 | **13.11** | 5.46 | 2.73 | 2.090 |
| IV | 42 | 2.19 | 14.75 | 6.56 | **20.77** | 1.09 | 16.39 | 2.147 |
| V | 34 | 16.94 | **27.87** | 2.73 | 2.19 | 0.00 | 0.00 | 1.426 |

| | | | | | | | |
|---|---|---|---|---|---|---|---|
| VI | 25 | **51.91** | 3.28 | 13.11 | 1.64 | 0.00 | 0.00 | 1.106 |
| VII | 18 | **43.17** | 0.55 | 12.57 | 0.00 | 0.00 | 30.05 | 1.481 |
| VIII | 42 | 0.00 | 0.55 | **37.70** | 2.19 | 8.74 | 1.64 | 1.182 |
| IX | 27 | **33.33** | 0.00 | 6.56 | 0.00 | 11.48 | 8.20 | 1.671 |
| X | 23 | 2.19 | 0.55 | 12.02 | **22.95** | 10.93 | 1.64 | 1.917 |

To overcome the limitations of nominal label-based categorizations, we analyzed the top 1 prediction using a dimensional semantic framework inspired by classical Rorschach constructs. Instead of assigning each ImageNet class to a single category, labels were mapped onto multiple latent dimensions reflecting perceptual and affective properties: *Threat*, *Animacy*, *Movement*, *Softness*, *Structural rigidity*, and *Natural form*. Each predicted class could activate one or more dimensions based on lexical and semantic cues.

**Sup. Table 2** reports, for each inkblot, the proportion of models whose dominant prediction activated each dimension, together with a Shannon entropy index summarizing the dispersion of the dimensional profile. This approach substantially reduces the amount of uninterpretable residual variance and reveals structured, non-random patterns across stimuli. We would like to remark that percentages reported in the table do not necessarily sum to 100%, as semantic dimensions are not mutually exclusive. Each predicted class may activate multiple properties, resulting in a multi-dimensional representation of model responses rather than a categorical partition.

Several inkblots exhibit a clear dominance of specific dimensions. Inkblots 6 and 7 are characterized by a strong *Threat* component (>40–50%), often co-occurring with *Movement*, indicating interpretations centered on dynamic and potentially aggressive objects. Inkblot 5 shows the highest *Animacy* score (≈28%), consistent with a biological organization of the percept. Inkblot 8 is dominated by *Movement*, while Inkblot 4 displays a more heterogeneous profile with elevated *Soft* and *Natural* dimensions, reflected in the highest dimensional entropy.

Lower entropy values are associated with stimuli eliciting stereotyped interpretations along a limited set of dimensions, whereas higher entropy values indicate a more diffuse and ambiguous perceptual organization. Importantly, these dimensional patterns emerge consistently across model architectures, suggesting that they are primarily driven by shared perceptual biases acquired during large-scale natural image training rather than by idiosyncratic architectural effects.

Overall, the dimensional analysis shows that responses produced by AI models to ambiguous Rorschach stimuli exhibit systematic patterns that can be mapped onto affective and perceptual dimensions commonly used in human Rorschach assessment, reflecting similarities at the level of representation rather than underlying psychological processes.

## Similarity analysis on model embedding space

We applied a dimensionality reduction on the AI model outputs to evaluate the differences between architecture responses (ref. **Figure 3** of the *Manuscript*). For each Rorschach image, we evaluated the features vectors in logit format produced by the AI models pre-trained on ImageNet dataset, as proxy of the internal decision space of the models before the application of the softmax activation function.

The logit vectors were standardized and normalized according to the respective z-score; we used a PaCMAP algorithm for the projection of the information into a bidimensional embedding space, monitoring the spread of points according to the different model families.

# Distance AI families and humans

To summarize differences across multiple psycho-semantic dimensions simultaneously, multivariate profiles were constructed by averaging standardized scores within each *model family*. Distances between model families and human references were quantified in this standardized space using Euclidean metrics, allowing the construction of a ranking of model families based on their multivariate similarity to human responses. The contribution of individual variables to these distances was further decomposed to identify which psycho-semantic dimensions most strongly differentiated models from humans.

All statistical analyses were conducted preserving the hierarchical structure of the data, and checks of robustness were performed where appropriate to avoid overly optimistic parametric assumptions.

**Sup Tab 3 | Ranking of model families by similarity to the human profile.** Model families ranked according to their multivariate distance from the human psycho-semantic profile, computed in standardized (z-score) space. Lower values indicate greater relative similarity to human responses. Distances ($d$) were computed as Euclidean distances between family-level mean profiles in standardized psycho-semantic space. The contributing variables correspond to those accounting for the largest proportion of the distance from the human reference profile.

| Rank | Model family | $d$ to Human | Main contributing variables |
|---|---|---|---|
| 1 | ConvNeXt | 10.36 | Dynamicity, Semantic complexity, Emotional valence |
| 2 | ShuffleNet | 10.44 | Dynamicity, Response frequency, Semantic complexity |
| 3 | RegNet | 10.44 | Dynamicity, Formal integration, Response frequency |
| 4 | EfficientNet | 10.48 | Response frequency, Semantic complexity, Dynamicity |
| 5 | DenseNet | 10.61 | Formal integration, Response frequency, Dynamicity |
| 6 | EfficientNetV2 | 10.64 | Response frequency, Formal integration, Dynamicity |
| 7 | VGG | 10.65 | Formal integration, Response frequency, Emotional valence |
| 8 | ResNet | 10.68 | Formal integration, Response frequency, Dynamicity |
| 9 | Inception / GoogLeNet | 10.83 | Response frequency, Semantic complexity, Dynamicity |
| 10 | MobileNet | 10.84 | Response frequency, Formal integration, Semantic complexity |
| 11 | SqueezeNet | 11.50 | Response frequency, Formal integration, Reduced dynamicity |

# Inter-model attentional agreement

To characterize how different deep learning models attend to ambiguous visual stimuli, we analyzed the Grad-CAM maps produced by 61 ImageNet-pretrained convolutional neural networks in response to the ten Rorschach inkblots. For each model and each inkblot, Grad-CAM maps were computed with respect to the top 1 predicted class and normalized to the $[0, 1]$ range.

First, we quantified inter-model attentional agreement at the level of each inkblot by computing pairwise similarities between Grad-CAM maps using multiple metrics capturing complementary aspects of spatial alignment: Pearson correlation (linear correspondence), cosine similarity (global directional agreement), structural similarity index (SSIM; local structural coherence), and overlap-based measures (Intersection-over-Union and Dice coefficient) computed on thresholded high-saliency regions. For each inkblot, these pairwise similarities were summarized by their mean off-diagonal value, yielding a global index of attentional convergence.

Second, the pairwise similarities were interpreted as weighted edges in a proximity network, where nodes represent models, and edges connect pairs of models whose attentional maps exceed a similarity threshold. On these networks, we performed community detection using greedy modularity maximization to identify groups of models sharing similar attentional strategies. The correspondence between detected communities and architectural model families was quantified using Normalized Mutual Information (NMI) and Adjusted Rand Index (ARI). In addition, a global network was constructed by averaging pairwise similarities across all inkblots, providing a stimulus-independent view of the organization of model attention.

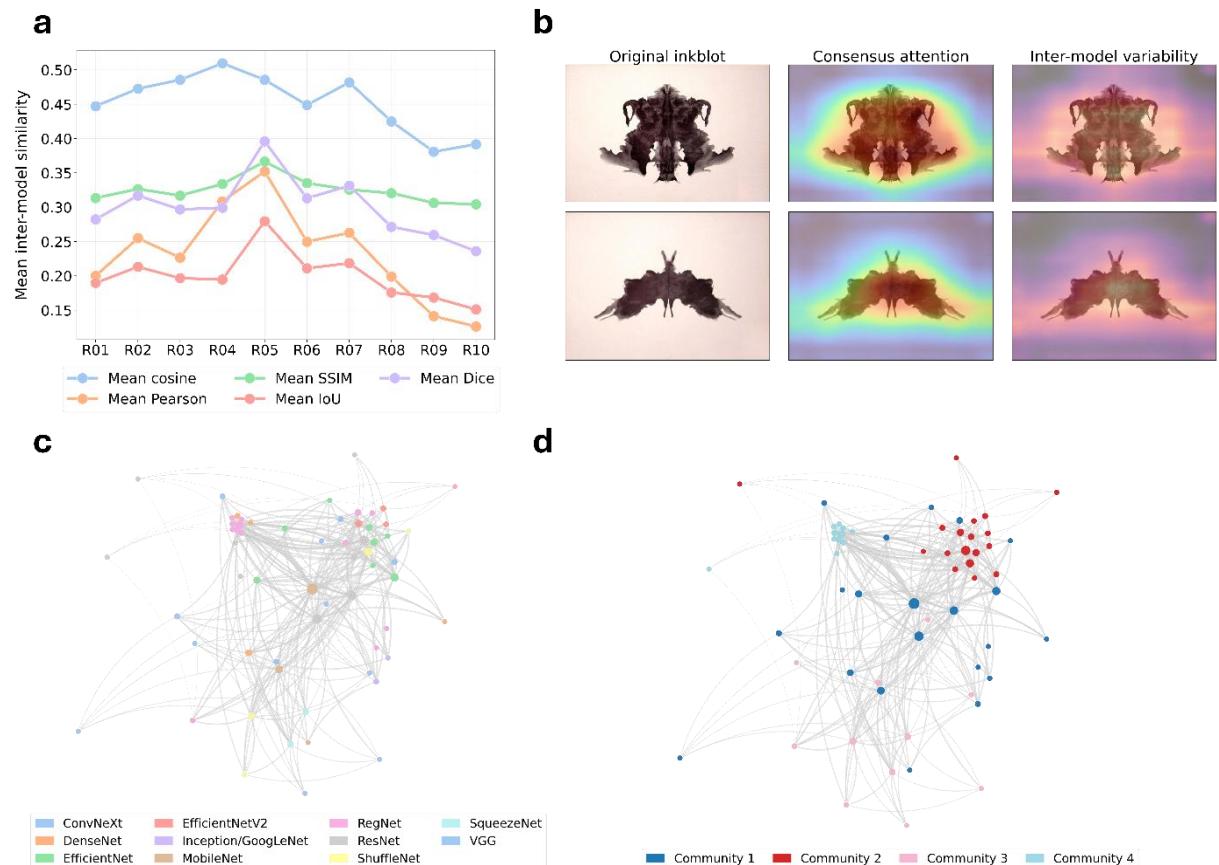

**Sup Figure 2 | Inter-model attentional agreement and network organization of Grad-CAM responses to Rorschach inkblots.** **a.** Mean inter-model agreement of Grad-CAM maps for each Rorschach inkblot, quantified using complementary similarity measures (cosine similarity, Pearson correlation, structural similarity index, Intersection-over-Union and Dice coefficient); agreement varies across inkblots, indicating stimulus-dependent convergence of model attention; **b.** for two representative inkblots, visualization of the original stimulus (left), the consensus attention map obtained by averaging normalized Grad-CAMs across all models (center), and the inter-model variability map computed as the pixel-wise standard deviation of Grad-CAMs (right); consensus maps highlight shared salient regions, whereas variability maps reveal spatial locations associated with divergent attentional strategies; **c.** global attentional proximity network with nodes representing models and edges encoding mean Grad-CAM similarity aggregated across inkblots; node colors indicate model families and node size is proportional to node degree; the network illustrates that models from different architectural families can occupy similar regions of the attentional space; **d.** same global proximity network as in (**c**), with node colors indicating communities identified via modularity-based community detection; communities reflect shared attentional strategies that cut across architectural families, revealing latent organizational structure in the models' perceptual behavior.

**Image-dependent attentional convergence**

The degree of attentional agreement between models varied substantially across inkblots. Some stimuli (e.g., *R04* and *R05*) elicited relatively high inter-model convergence, with mean cosine similarities approaching 0.5 and comparatively higher Dice and IoU values. Other inkblots (notably *R09* and *R10*) showed markedly lower agreement across all metrics.

Across all stimuli, cosine similarity values were consistently higher than Pearson correlations, indicating that models tend to agree on the global spatial distribution of salient regions while differing in local intensity patterns and fine-grained structure. The moderate overlap-based scores further suggest that convergence occurs at the level of functionally relevant regions rather than exact pixel-wise correspondence.

An analogous approach was not performed on human evaluation due to missing information and difficulties in collecting a detailed description of attention maps. Future analysis will take care of this comparison after an accurate data collection.

**Consensus maps and model centrality**

By averaging normalized Grad-CAM maps across models, we derived a consensus attentional map for each inkblot. The similarity of individual models to this consensus revealed systematic architectural differences. EfficientNet, Inception/GoogLeNet, and MobileNet families showed the highest similarity to the consensus, whereas VGG and, to a lesser extent, ResNet models were aligned more weakly.

At the level of individual architectures, several relatively lightweight or shallow models (e.g., MobileNetv2, ResNet18) emerged as highly "central," closely approximating the consensus despite their lower parameter count. This indicates that proximity to the collective attentional pattern is not trivially explained by model size or depth but rather reflects how efficiently a model captures broadly shared visual cues.

**Network structure and community organization**

Representing pairwise Grad-CAM similarities as networks revealed a rich relational structure that is not apparent from mean agreement values alone. Inkblots with higher attentional convergence produced denser networks with higher average degree and fewer connected components, whereas more ambiguous stimuli yielded fragmented graphs with multiple disconnected components, reflecting divergent attentional strategies.

Community detection consistently identified a relatively large number of communities (≈10–17 per inkblot), exceeding the number of architectural families. The correspondence between attentional communities and model families was moderate when assessed with NMI, but weak or negligible according to ARI. This dissociation indicates that while architectural lineage contributes to shaping attentional behavior, it does not uniquely determine it. Models from different families can converge onto similar attentional strategies, and conversely, models from the same family can diverge substantially.

The global proximity network, aggregating similarities across all inkblots, reinforced this conclusion. The presence of many communities and a moderate NMI but near-zero ARI suggests the existence of latent attentional modes that persist across stimuli yet cut across architectural boundaries.

**Attention convergence interpretation**

Taken together, these analyses show that deep neural networks exhibit structured but non-trivial patterns of attentional convergence when confronted with perceptually ambiguous stimuli such as Rorschach inkblots. The models do not respond randomly: instead, their attention converges partially in an image-dependent manner, giving rise to shared saliency patterns, consensus maps, and stable relational structures.

At the same time, this convergence does not collapse into a single dominant strategy, nor is it rigidly dictated by architectural families. Rather, the results point to a distributed organization of perceptual strategies, shaped by the statistics of natural images learned during training and modulated by the intrinsic ambiguity of the stimulus.

In this sense, the Rorschach test functions as a probe of the latent perceptual geometry of deep vision models, revealing how shared training experience and architectural constraints jointly give rise to convergent yet plural interpretations of stimuli that, by design, lack a single correct meaning.

# Human responses analysis

Data on human responses to the Rorschach test were extracted from established normative sources reported in classical psychodiagnostics literature. We used a collection of standardized responses obtained from non-clinical samples, in which interpretations of the Rorschach inkblots were encoded according to accepted classification systems. This information was used as a statistical reference, without diagnostic or clinical intent.

## Dataset structure

For each Rorschach picture, the human-dataset includes a set of popular responses, defined as those interpretations that exceed a minimum frequency threshold in the reference population. Each response was associated with an absolute frequency and/or percentage, derived from the number of subjects who produced that specific interpretation. Responses were stored in normalized text form (lemma or summary description), allowing for subsequent computational processing. Formally, for each picture $j$, the human dataset was described as pairs $(r_{jt}, w_{jt})$ where $r_{jt}$ is the textual response and $w_{jt}$ represents the associated frequency (absolute or percentage).

## Human responses data

The following table reports the complete set of popular human responses included in the analyses, organized by Rorschach inkblot. Percentages indicate the frequency with which each response is reported in the normative reference data. These responses constitute the input dataset for the computation of psycho-semantic indices and for the comparison with AI model predictions described in the main text. We would like to remark that percentages reported in the table do not necessarily sum up to 100%, as the same subject could provide multiple interpretations and there is not a complete agreement among popular responses in literature.

**Sup Tab 4 | Popular human responses to the Rorschach test used in the analysis.** The table reports the set of popular human responses for each Rorschach inkblot, together with their associated normative frequencies (expressed as percentages). Responses were extracted from consolidated normative sources in the Rorschach literature and represent interpretations that recur most frequently in non-clinical populations. The selection and labeling of responses are consistent with standard Rorschach coding systems, including classical manuals and contemporary systems such as Exner's Comprehensive System and the Rorschach Performance Assessment System (R-PAS). These data were used exclusively as aggregated statistical references to enable a quantitative comparison with artificial intelligence model predictions and were not employed for diagnostic or clinical purposes.

| Inkblot | Domain | Response (description) | Frequency (%) |
|---|---|---|---|
| I | Animal | Bat | 45 |
| I | Animal | Butterfly | 40 |
| I | Animal | Moth | 22 |
| I | Animal | Bird | 18 |
| I | Form | Symmetrical figure | 25 |
| II | Human | Two people arguing | 30 |
| II | Body | Blood | 35 |
| II | Animal | Two animals fighting | 15 |
| II | Fantasy | Monster | 10 |
| II | Abstract | Red blot | 22 |
| III | Human | Two people talking | 50 |
| III | Human | Two people cooperating | 42 |
| III | Human | Waiters | 35 |
| III | Relationship | Social interaction | 40 |
| III | Object | Masks | 10 |
| IV | Human | Large man | 55 |
| IV | Human | Authoritative figure | 40 |
| IV | Animal | Gorilla | 22 |
| IV | Fantasy | Monster | 15 |
| IV | Object | Statue | 8 |
| V | Animal | Bat | 60 |
| V | Animal | Butterfly | 50 |
| V | Form | Symmetrical figure | 35 |
| V | Form | Cross | 12 |
| V | Fantasy | Winged creature | 6 |
| VI | Object | Animal skin | 45 |
| VI | Object | Fur | 38 |
| VI | Body | Skin | 30 |
| VI | Sensation | Soft object | 25 |
| VI | Fantasy | Animal remains | 10 |
| VII | Human | Two women | 45 |
| VII | Human | Mother and child | 28 |
| VII | Relationship | Affective scene | 20 |
| VII | Fantasy | Ghosts | 12 |
| VII | Object | Masks | 10 |
| VIII | Animal | Animals in motion | 35 |
| VIII | Animal | Marine creatures | 22 |
| VIII | Landscape | Natural scene | 25 |
| VIII | Fantasy | Fantastic animals | 15 |
| VIII | Abstract | Colored composition | 30 |
| IX | Abstract | Indefinite shape | 40 |
| IX | Event | Explosion | 25 |
| IX | Event | Smoke | 30 |
| IX | Fantasy | Monster | 15 |
| IX | Process | Chemical reaction | 10 |
| X | Animal | Insects | 35 |
| X | Animal | Marine creatures | 30 |
| X | Geography | Map | 20 |
| X | Chaos | Confused scene | 40 |
| X | Fantasy | Multiple creatures | 25 |

## Response pre-processing

The human text responses were pre-processed according to a linguistic normalization phase, removing non-informative lexical variability. The developed pre-processing involved: (i) conversion to lemma form; (ii) removal of non-semantic articles and modifiers; and (iii) unification of obvious synonyms (e.g., singular/plural, spelling variants). No transformations were performed that could alter the semantic content of the responses. This preprocessing is intended to facilitate alignment with ImageNet labels in subsequent analyses.

## Semantic mapping and categorization

Each human response was projected into a semantic vector space using text embeddings to guarantee a fair comparison with the outcomes of AI models. Semantically equivalent or highly similar responses were aggregated into conceptual categories based on a predefined similarity threshold. This step allows us to treat lexically different but conceptually overlapping human responses as equivalent.

## Psycho-semantic attribution

Each human response was associated with psycho-semantic attributes used in comparative analyses, including emotional valence, affective load, human reference, social orientation, aggression, anxiety, dynamism, and semantic complexity. These attributes were derived from coding methods reported in literature or from predefined operational rubrics. The coding scales and item meanings are described in the next sections.

The analyses were performed aggregating human responses according to Rorschach inkblots. For each picture, the distributions of human responses were used to calculate summary indices (percentage frequencies, weighted averages of psycho-semantic indices, semantic dispersion), ensuring direct comparability with the measures extracted from AI predictions.

# Extended analysis of model responses

To characterize how artificial neural networks respond to highly ambiguous visual stimuli, we adopted a multi-level analytical strategy that progressively incorporates increasing semantic structure into the interpretation of model predictions. Starting from a purely symbolic description of predicted labels, we then integrated hierarchical semantic information from WordNet and finally examined the conceptual proximity of predictions using distributional semantic embeddings. This section summarizes the analyses performed and the main quantitative findings obtained at each level.

## Label-Based Analysis of Predicted Classes

In the first stage of the analysis, predicted classes were treated as nominal labels, without any explicit consideration of their semantic relationships. For each image–model pair, the top 1 predicted class was extracted, yielding 610 observations (10 images × 61 models). The goal of this analysis was to determine whether the distribution of predicted labels depended on *model identity*, *model family*, or *model size*.

Contingency analyses revealed a strong association between predicted class and *model family* ($\chi^2$ = 1310.64, p = 6.1 × 10$^{-8}$), with a Cramér's V of 0.46, indicating a moderate-to-large effect size. In contrast, the association between predicted class and individual *model identity* did not reach statistical significance after accounting for the large number of degrees of freedom ($\chi^2$ = 6427.69, p = 0.13), despite a non-negligible Cramér's V (0.42). This pattern suggests that prediction behavior clusters primarily at the level of *architectural families* rather than individual model instances.

*Model size* also showed a significant association with predicted class when discretized into quartiles ($\chi^2$ = 399.47, p = 8.7 × 10$^{-4}$; Cramér's V = 0.47), an effect that remained significant after false discovery rate (FDR) correction (p = 1.3 × 10$^{-3}$). Together, these results demonstrate that even in the absence of ground-truth labels, the choice of class is systematically influenced by architectural design and model capacity, rather than being random or idiosyncratic.

**Sup Tab 5 | Label-based association between predicted classes and model characteristics.** Results of contingency analyses testing the association between top 1 predicted classes and model characteristics. Reported are chi-square statistics ($\chi^2$), degrees of freedom (df), raw p-values, FDR-corrected p-values, and Cramér's V as a measure of effect size.

| Predictor | $\chi^2$ | df | p-value | p-FDR | Cramér's V |
|---|---|---|---|---|---|
| Model family | 1310.64 | 1050 | 6.13 × 10$^{-8}$ | 1.84 × 10$^{-7}$ | 0.46 |
| Individual model | 6427.69 | 6300 | 0.128 | 0.128 | 0.42 |
| Number of parameters (quartiles) | 399.47 | 36 | $8.73 \times 10^{-4}$ | $1.31 \times 10^{-3}$ | 0.47 |

## WordNet-Based Semantic Analysis

To move beyond symbolic labels and incorporate semantic structure, predicted classes were mapped to *WordNet synsets*, enabling analysis at the level of semantic domains and hierarchical relations. Nearly all predictions (3049 out of 3050 top 5 predictions) could be successfully mapped to *WordNet synsets*, confirming that most predicted categories correspond to well-defined concepts.

At the top 1 level, predictions were strongly dominated by a small number of semantic domains. Artifacts accounted for 387 out of 610 predictions, followed by animals (126), with smaller contributions from objects, plants, and food categories. Statistical analysis confirmed that the semantic domain (*WordNet lexname*) of the predicted class depended both on *model family* ($\chi^2$ = 148.54, p = 0.040, Cramér's V = 0.16) and *model size* ($\chi^2$ = 58.32, p = 0.011, Cramér's V = 0.18). Although these effect sizes are modest, both associations remained significant after FDR correction, indicating reproducible semantic biases across architectures and capacity levels.

Extending the analysis to the top 5 predictions, we examined semantic properties of the candidate sets. Mixed-effects models revealed that some families tended to propose semantically more specific concepts, as reflected by a significantly greater average WordNet depth of the top 5 classes ($\beta$ = 1.56, p = $2.1 \times 10^{-4}$). However, no robust differences emerged in terms of semantic diversity within the top 5, as measured by WordNet-based similarity metrics: after FDR correction, no model family or size effect remained significant. Thus, while architecture differs in the type and specificity of concepts they propose, they do not differ substantially in the overall semantic coherence of their predictions.

Sup Tab 6 | WordNet-based semantic domain (*lexname*) analysis of top 1 predictions. Association between WordNet semantic domains (*lexnames*) of top 1 predicted classes and model characteristics. Semantic domains reflect high-level conceptual categories such as artifacts, animals, and food.

| Predictor | $\chi^2$ | df | p-value | p-FDR | Cramér's V |
|---|---|---|---|---|---|
| Model family | 148.54 | 120 | $3.96 \times 10^{-2}$ | $3.96 \times 10^{-2}$ | 0.16 |
| Number of parameters (quartiles) | 58.32 | 36 | $1.07 \times 10^{-2}$ | $2.14 \times 10^{-2}$ | 0.18 |

## Conceptual Proximity Analysis Using Distributional Embeddings

Finally, to capture a notion of conceptual similarity that is not restricted to hierarchical taxonomies, we embedded all predicted class labels into a distributional semantic space using pre-trained word vectors. This approach allows distances between classes to reflect conceptual relatedness as learned from natural language usage. All 258 unique class labels were successfully embedded into a 300-dimensional space.

For each image–model pair, two concept-level metrics were computed on the top 5 predictions: the entropy of the probability distribution and the conceptual spread, defined as the average cosine distance between class embeddings. Mixed-effects models were again used to assess the influence of model family and size, with images treated as a random effect.

The analysis revealed a clear architectural signature in probabilistic uncertainty. DenseNet, SqueezeNet, and VGG architectures exhibited significantly higher top 5 entropy compared to the reference family (DenseNet: $\beta$ = 0.176, p = $4.7 \times 10^{-4}$; SqueezeNet: $\beta$ = 0.218, p = $4.7 \times 10^{-4}$; VGG: $\beta$ = 0.141, p = $6.6 \times 10^{-4}$). These coefficients indicate a systematically flatter probability distribution over the top 5 classes, corresponding to greater uncertainty or a broader set of competing hypotheses. The effect of model size on entropy was weak (β = 0.024, p = 0.046) and did not survive FDR correction (p = 0.16).

In contrast, conceptual spread showed no statistically robust dependence on either model family or model size after correction for multiple comparisons. Although EfficientNet and ResNet displayed slightly reduced spread at the level of uncorrected p-values (p ≈ 0.03), these effects were not stable under FDR correction. Thus, while some architectures distribute probability mass more broadly across multiple classes, the candidate classes themselves remain conceptually close to one another across all architectures.

The variance associated with the image-level random effect was consistently small for both metrics, indicating that differences between models dominate over differences between images in shaping the conceptual structure of the predictions for the specific stimulus set considered.

**Sup Tab 7 | Mixed-effects models on WordNet-based semantic metrics (top 5 predictions).** Summary of linear mixed-effects models assessing the influence of model family and size on WordNet-based semantic properties of the top 5 predictions. Reported are regression coefficients (β) for significant fixed effects only, with FDR-corrected p-values. Only SqueezeNet shows a robust tendency to propose more fine-grained semantic concepts; no architecture exhibits a reliable difference in semantic diversity after correction.

| Dependent variable | Significant predictor | β (estimate) | p-FDR |
|---|---|---|---|
| Average semantic depth (probability-weighted) | SqueezeNet family | +1.56 | $2.11 \times 10^{-4}$ |
| Semantic diversity (Wu–Palmer) | — | — | n.s. |

**Sup Tab 8 | Conceptual proximity analysis using distributional embeddings.** Results of mixed-effects models assessing conceptual organization of top 5 predictions in a distributional semantic space. Probabilistic entropy quantifies uncertainty across candidate classes, while conceptual spread measures average conceptual distance among them.

| Dependent variable | Predictor | β (estimate) | p-FDR |
|---|---|---|---|
| Probabilistic entropy | DenseNet family | +0.176 | $4.70 \times 10^{-4}$ |
| | SqueezeNet family | +0.218 | $4.70 \times 10^{-4}$ |
| | VGG family | +0.141 | $6.63 \times 10^{-4}$ |
| | Number of parameters | +0.024 | n.s. |
| Conceptual spread | Any predictor | — | n.s. |

## Psycho-semantic metrics

To enable a quantitative comparison between human Rorschach responses and artificial image classification outputs, we introduced a set of psycho-semantic descriptors associated with each response label. These descriptors are stimulus-independent, non-diagnostic, and rule-based, and aim to capture affective, perceptual, and semantic properties of response contents rather than subject-level psychological traits. All metrics are defined at the response level and can therefore be applied uniformly to human verbalizations, AI-generated class labels, and intermediate semantic representations.

For each response $r$, we define a vector of psycho-semantic features:
$$\mathbf{p}(r) = [V(r), A(r), P(r), I(r), S(r), H(r), G(r), T(r), C(r), D(r)]$$
where each component is defined below.

## Affective dimensions

### Emotional valence

The emotional valence $V(r)$ captures the intrinsic affective polarity of the semantic content of the response.
$$V(r) \in \{-2, -1, 0, +1, +2\}$$
where values equal to +2 identifies strongly positive values (pleasant, rewarding, affiliate), +1 identifies moderately positive values, 0 identifies neutral/descriptive values, -1 identifies moderately negative values, and -2 identifies strongly negative values (threating, violent, aversive). Assignment is based on semantic category (e.g., *food*, *animals*, *weapons*), independent of the stimulus context.

**Affective load**
The affective load $A(r)$ quantifies whether the response carries emotional salience beyond neutral description:
$$A(r) = \begin{cases} 1 & \text{if } V(r) \neq 0 \\ 0 & \text{if } V(r) = 0 \end{cases}$$
This conservative definition avoids over-weighting affective interpretations and ensures binary interpretability.

## Threat- and conflict-related dimensions

**Aggressiveness**
The aggressiveness index $G(r)$ reflects explicit or implicit violent or offensive content:
$$G(r) \in \{0, 1, 2, 3\}$$
where 0 identifies 'Not aggressive content', 1 identifies potentially mild content (e.g. harmful animals), 2 identifies moderate content (e.g. explicit weapons or conflict), and 3 identifies a high content (e.g. warfare, lethal weapons).

**Anxiety potential**
The anxiety index $T(r)$ estimates the potential of the response content to elicit fear, danger, or aversion:
$$T(r) \in \{0, 1, 2, 3\}$$
High values are assigned to responses involving weapons, violent scenarios, dangerous animals, and catastrophic natural phenomena.

## Perceptual-adaptive dimensions

**Perceptual adaptivity**
The perceptual adaptivity $P(r)$ represents the degree of semantic conventionality of the response:
$$P(r) \in [0, 1]$$
Higher values correspond to highly conventional, well-defined semantic objects. In the stimulus-independent annotation used here, $P(r)$ was fixed to a baseline value to avoid introducing frequency-based bias outside normative Rorschach contexts.

**Formal integration**
The formal integration index $I(r)$ measures the internal semantic coherence of the response:
$$I(r) \in \{0, 1, 2\}$$
where 0 indicates vague or poorly structured coherence, 1 a moderately structured coherence, and 2 a well-defined and semantically coherent response. All object-like ImageNet labels were assigned to the maximal value.

## Social and human-related dimensions

### Social orientation

The social orientation $S(r)$ captures whether the semantic content implies interpersonal interaction:
$$S(r) \in \{-1, 0, +1\}$$
where +1 is associated with prosocial/cooperative orientation, 0 is associated with a neutral orientation, and -1 is associated with a conflictual/antisocial orientation. In the considered dataset, this metric was conservatively set to zero unless explicit interaction was encoded in the label.

### Human presence

The human presence index $H(r)$ quantifies explicit or implicit reference to the human body or human agents:
$$H(r) \in \{0, 1, 2\}$$
where 0 indicates the absence of references, 1 indicates an implicit presence (e.g. clothing, tools), and 2 indicates an explicit presence (e.g. human figures or body parts).

## Cognitive-semantic complexity

### Semantic complexity

The semantic complexity $C(r)$ reflects the structural richness of the response:
$$C(r) \in \{1, 2, 3\}$$
where 1 indicates a single object, 2 indicates a relation between objects, and 3 indicates a scene or process.

### Dynamism

The dynamism index $D(r)$ captures implied motion or action:
$$D(r) \in \{0, 1, 2\}$$
where 0 indicates a static action, 1 indicates an implicit motion, and 2 indicates an explicit action. Animals, vehicles, and weapons were assigned higher dynamism values than static objects.